\documentclass[prd,aps,showpacs,nofootinbib]{revtex4-1}
\usepackage{amssymb}
\usepackage{amsfonts}
\usepackage{amsmath}
\usepackage{graphicx}

\newcommand{\be}{\begin{equation}}
\newcommand{\ee}{\end{equation}}
\newcommand{\bq}{\begin{eqnarray}}
\newcommand{\eq}{\end{eqnarray}}

\begin{document}

\title{Connection between Symmetrical Special Relativity and the Gravitational Bose Einstein Condensate of a Gravastar/Dark Energy
Star: Are there singularities in spacetime like black holes?}
\author{*Cl\'audio Nassif and **Rodrigo Francisco do Santos}

\affiliation{\small{{\bf*CPFT}:{\bf Centro de Pesquisas em F\'isica Te\'orica}, Rua Rio de Janeiro 1186, Lourdes, CEP:30.160-041,
 Belo Horizonte-MG, Brazil.\\
 {\bf **UFF:Universidade Federal Fluminense}, Av.Litoranea s/n, Gragoata, CEP:24210-340, Niteroi-RJ, Brazil.\\
 cnassifcbpf@yahoo.com.br; santosst1@gmail.com}} 

\begin{abstract}
We aim to search for a connection between an invariant minimum speed that breaks down the Lorentz symmetry and the Gravitational 
Bose Einstein Condensate (GBEC), which is the central core of a star of gravitating vacuum (Gravastar/Dark Energy Star) by introducing a 
cosmological constant into compact objects. This model was designed to circumvent the embarrassment generated by the paradoxes of a 
singularity as the final stage of a gravitational collapse, by introducing in place of the
singularity of event horizon a spatial-temporal phase transition, a concept with which the causal structure of Symmetrical Special 
Relativity (SSR) helps us to elucidate by providing a quantum interpretation for GBEC and explaining the origin of anisotropy, which has 
been introduced in ad-hoc way before in the literature.
\end{abstract}


\maketitle

\section{Introduction}

The search for understanding the cosmological vacuum has been the issue of hard investigations\cite{1}. Several models have been
suggested\cite{2} in order to postulate the existence of a quantum vacuum (e.g: fluid of Zeldovich), but without proposing
a physical interpretation for the vacuum until the emergence of Symmetrical Special Relativitty (SSR)\cite{3}\cite{4}\cite{5}\cite{6},
which has brought a new interpretation for the quantum vacuum by means of the concept of an invariant minimum speed,
which changes the causal structure of spacetime and geometrizes the quantum phenomena\cite{5}. This minimum speed modifies the basic
metric of spacetime by provinding a metric of type De-Sitter (DS). Such a metric describes the geometric structure of the
Gravitational Bose Einstein Condensate (GBEC), which represents the core of a gravastar\cite{7}\cite{8}\cite{9}\cite{10}\cite{16}\cite{17}
\cite{18}\cite{19}\cite{20}. So we are able to map them in such a way that we can associate the cosmological constant with the minimum
speed connected to the vacuum energy density\cite{3}. We still associate the star's compactness with a factor that include the minimum 
speed in a type of DS-metric. Thus by linking the star structure equations with the causal structure of spacetime of SSR, a phase 
transition appears in place of the horizon of event as postulated by Chapline and other\cite{11}\cite{12}\cite{14}\cite{15}. Such a phase
transition can be related to this new causal structure of SSR, as Nassif has demonstrated in ref.\cite{3}, where SSR describes perfectly
a fluid which is similar to a relativistic superfluid of type of cosmological fluid, being the constituent of GBEC.

In the 2nd. section, we introduce the concept of a minimum speed given in the intergalactic spacetime with the presence of a weak gravitational
field\cite{3}. Thus we present the spacetime metric with such a new causal structure and so we show the cosmological fluid 
generated by SSR.

In the 3rd. section, we introduce the GBEC. We make a brief study about the concept of phase transition that appears in place of
the horizon of event for the gravitational collapse.

In section 4, we discuss the relationship between the concepts of SSR and GBEC, where we apply the concepts of SSR to GBEC of
a Gravastar/Dark Energy Star by relating DS-metric with SSR metric. We write the Tolman-Oppenheimer-Volkof (TOV) equation and the
conservation of mass in terms of the causal structure of SSR. We also introduce the typical quantum concepts of SSR that appear naturally
in TOV equation by showing how difficult is to penetrate the core of a gravastar (GBEC). Thus, we associate the vacuum structure generated 
by an anisotropy with the phase transition (gravity/anti-gravity of the GBEC), which appears in place of the horizon of event by preventing 
the emergence of singularities. 

In section 5, we investigate deeper what occurs in the region of phase transition, i.e., at the coexistence radius of the two phases. We
aim to obtain the spacetime metric in such a region where we verify that the metric does not diverge as occurs in the classical case 
for the Schwarzchild radius (no phase transition). We show that there is no divergence at such a transition region due to the fact that
the coexistence radius of phases is slightly larger than the Schwarzchild radius by preventing the singularity. 

\section{The causal structure of SSR}

A breakdown of the Lorentz symmetry for very low energies\cite{3}\cite{4}\cite{5}\cite{6} generated by the presence of a
residual (weak) background gravitational field creates a new causal structure in spacetime, where we have a mimimum speed $V$ that is
unattainable for all particles, and also a universal dimensionless constant $\xi$\cite{3}, which couples the gravitational field
to the electromagnetic one, namely:

\begin{equation}
\xi=\frac{V}{c}=\sqrt{\frac{Gm_{p}m_{e}}{4\pi}}\frac{q_{e}}{\hbar c},
\end{equation}
where $m_{p}$ and $m_{e}$ are respectively the mass of the proton and electron. The value of such a minimum speed is
$V=4.5876\times 10^{-14}$ m/s. We find $\xi=1.5302\times 10^{-22}$. 

It was shown\cite{3} that the minimum speed is connected to the cosmological constant in the following way:

\begin{equation}
 V\approx\sqrt{\frac{e^{2}}{m_{p}}\Lambda^{\frac{1}{2}}}
\end{equation}.

Therefore the light cone contains a new region of causality called {\it dark cone}\cite{3}, so that the velocities of the particles
must belong to the following range: $V$(dark cone)$<v<c$ (light cone) (Fig.1). 

The breaking of the Lorentz symmetry group destroys the properties of the transformations of Special Relativity (SR) and so generates
intriguing kinematics and dynamics for speeds very close to the minimum speed $V$, i.e., for $v\rightarrow V$, we find new 
relativistic effects such as the contraction of the improper time and the dilation of space\cite{3}. In this new scenario, the proper time
also suffers relativistic effects such as its own dilation with regard to the improper one when $v\rightarrow V$\cite{3}\cite{4}, namely:

\begin{equation}
\Delta\tau\sqrt{1-\frac{V^{2}}{v^{2}}}=\Delta t\sqrt{1-\frac{v^{2}}{c^{2}}}.
\end{equation}

Since the minimum speed $V$ is an invariant quantity as the speed of light $c$, so $V$ does not alter the value of the speed $v$ of
any particle. Therefore we have called ultra-referential $S_{V}$\cite{3}\cite{4} as being the preferred (background) reference frame in
relation to which we have the speeds $v$ of any particle. In view of this, the reference frame transformations change substantially
in the presence of $S_V$, as follow below:

\begin{figure}
\begin{center}
\includegraphics[scale=0.80]{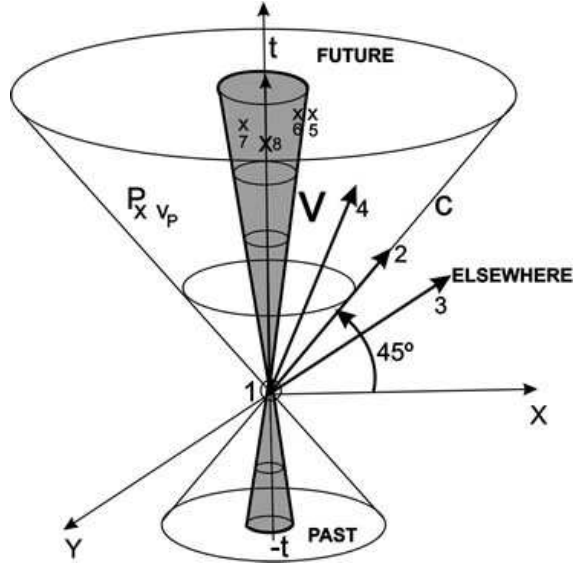}
\end{center}
\caption{The external and internal conical surfaces represent respectively the speed of light $c$ and the unattainable minimum
speed $V$, which is a definitely prohibited boundary for any particle. For a point $P$ in the world line of a particle, in the interior
of the two conical surfaces, we obtain a corresponding internal conical surface, such that we must have $V<v_p\leq c$. The $4$-interval
$S_4$ is of type time-like. The $4$-interval $S_2$ is a light-like interval (surface of the light cone). 
The $4$-interval $S_3$ is of type space-like (elsewhere). The novelty in spacetime of SSR are the $4$-intervals $S_5$ (surface of
the dark cone) representing an infinitly dilated time-like interval\cite{3}, including the $4$-intervals $S_6$, $S_7$ and 
$S_8$ inside the dark cone for representing a new region of type space-like\cite{3}.}
\end{figure}

The $(3+1)D$ transformations in SSR are

\begin{equation}
t'=\theta\gamma[t-\frac{\vec{r}\cdotp\vec{v}}{c^{2}}+\frac{\vec{r}\cdotp\vec{V}}{c^{2}}],
\end{equation}

where $\theta=\sqrt{1-\frac{V^{2}}{v^{2}}}$ and
$\theta\gamma=\psi=\frac{\sqrt{1-\frac{V^{2}}{v^{2}}}}{\sqrt{1-\frac{v^{2}}{c^{2}}}}$.

\begin{equation}
 \vec{r'}=\theta[\vec{r_{T}}+\gamma(\vec{r_{//}}-\vec{v}(1-\frac{V}{v}))t]=\theta[\vec{r_{T}}+\gamma(\vec{r_{//}}-\vec{v}t+\vec{V}t)]
\end{equation}

In the special case $(1+1)D$ with $\vec v=v_x=v$, we find the following transformations: $x^{\prime}=\psi(x-vt+Vt)$ and
$t^{\prime}=\psi(t-vx/c^2+Vx/c^2)$. The inverse transformations for the general case $(3+1)D$ and $(1+1)D$ were demonstrated in
ref.\cite{3}. Of course, if we make $V\rightarrow 0$, we recover the well-known Lorentz transformations.

Although we associate the minimum speed $V$ with the ultra-referential $S_{V}$, this is inaccessible for any particle. Thus, the effect
of such causal structure generates a symmetric mass-energy effect to what happens close to the speed of light $c$, i.e., it was shown
that $E=m\psi c^2$, so that $E\rightarrow 0$ when $v\rightarrow V$\cite{3}\cite{4}. It was also shown that the minimum speed $V$
is associated with the cosmological constant, which is equivalent to a fluid (vacuum energy) with negative pressure\cite{3}\cite{4}.

The metric of such symmetrical spacetime of SSR is a deformed Minkowisky metric with a global multiplicative function $\Theta$, being fully
equivalent to a DS-metric, namely:

\begin{equation}
ds^{2}={\Theta}g_{\mu\nu}dx^{\mu}dx^{\nu},
\end{equation}
where $\Theta=\theta^{-2}=1/(1-V^2/v^2)$\cite{3}\cite{4}.

We can say that SSR geometrizes the quantum phenomena as investigated before (the Uncertainty Principle)\cite{5} in order to allow us 
to associate quantities belonging to the microscopic world with a new geometric structure that originates from a Lorentz symmetry breaking.
Thus SSR may be a candidate to try to solve the problems associated with the gravitational collapse, which is a phenomenon that mixes 
inevitably quantum quantities with geometric structure of spacetime.

\subsection{The momentum-energy tensor in the presence of the ultra-referential $S_V$}

In this section we present the energy-momentum tensor\cite{3} which describes a relativistic superfluid. We write the
quadri-velocity of the superfluid, namely:

\begin{equation}
U^{\mu}=[U^{0},U^{\alpha}]=\left[\frac{\sqrt{1-\frac{V^2}{v^{2}}}}{\sqrt{1-\frac{v^2}{c^2}}},
\frac{v_{\alpha}}{c}\frac{\sqrt{1-\frac{V^2}{v^{2}}}}{\sqrt{1-\frac{v^2}{c^2}}}\right],
\end{equation}
where $\mu=0,1,2,3$ and $\alpha=1,2,3$. If we make $V\rightarrow 0$, we recover the quadri-velocity of SR. Now we write the
energy-momentum tensor, as follows:

\begin{equation}
T^{\mu\nu}=(p+\rho)U^{\mu}U^{\nu}-pg^{\mu\nu}.
\end{equation}

The signature of the tensor is $T^{\mu\nu}=Diag(\rho,p,p,p)$. So the component $T^{00}$ is

\begin{equation}
T^{00}=\frac{\rho(1-\frac{V^2}{v^{2}})+p(\frac{v^2}{c^2}-\frac{V^2}{v^2})}{\left(1-\frac{v^2}{c^2}\right)}.
\end{equation}

In the limit $V\rightarrow 0$, we recover SR. In the absence of sources, we have $\lim_{v\rightarrow V}=-pg^{\mu\nu}$, where we find
$\rho=-p$, which is the limiting case of the equation of state (EOS) associated with the cosmological vacuum. For a generic vacuum, we have
$p=-\frac{v^2}{c^2}\rho$. Thus SSR is able to describe in detail a superfluid which is very similar to what we see in the literature
on Gravastar/Dark Energy Star.

In a future work, we will explore fully this superfluid in various scenarios including gravitational collapse. Now, let us
investigate the structure of a gravastar.

\section{GBEC and the phase transition}

\begin{figure}
\begin{center}
\includegraphics[scale=0.80]{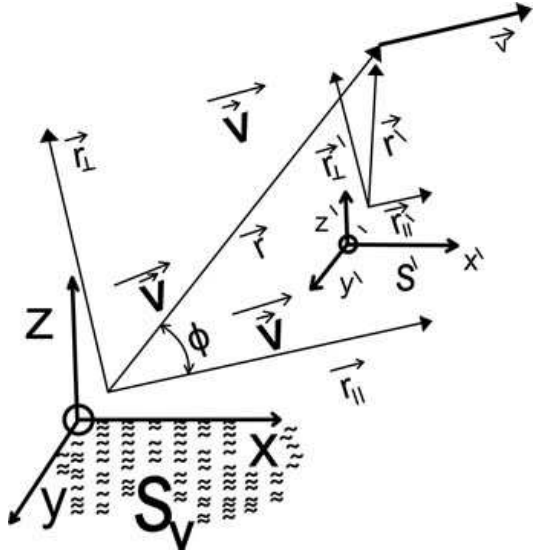}
\end{center}
\caption{$S^{\prime}$ moves with a $3D$-velocity $\vec v=(v_x,v_y,v_z)$ in relation to $S_V$. For the special case of $1D$-velocity
$\vec v=(v_x)$, we recover the case $(1+1)D$; however, in this general case of $3D$-velocity $\vec v$, there must be a background vector
 $\vec V$ (minimum velocity)\cite{3} with the same direction of $\vec v$ as shown in this figure. Such a background vector $\vec V=(V/v)\vec v$ is
related to the background reference frame (ultra-referential) $S_V$, leading to Lorentz violation. The modulus of $\vec V$ is invariant
at any direction.}
\end{figure}

The core of a Gravastar/Dark Energy Star is described as being composed of an exotic material called Gravitational Bose
Einstein Condensate (GBEC)\cite{9}\cite{16}\cite{17}. This is a relativistic superfluid as reported in the previous section.
This region connects with a shell of ordinary matter (baryonic matter) described by the Schwarzschild metric. Such a connection would
take place by means of a phase transition in spacetime\cite{12}\cite{14}\cite{15} that occurs near the Schwarzschild radius. Thus,
by following the works of CFV and MM\cite{9}\cite{10}\cite{13}\cite{16}\cite{17}\cite{24}, we write the metric of a gravastar
\cite{16}\cite{17}, namely:

\begin{equation}
ds^{2}=-f(r)c^{2}dt^{2}+\frac{dr^{2}}{h(r)}+r^2d\Omega^{2},
\end{equation}
where the metric functions $f(r)$ and $h(r)$ are given for DS-sector (GBEC), namely:

\begin{equation}
f_{GBEC}(r)=C\left(1-\frac{r^2}{R^2_{core}}\right)
\end{equation}
and
\begin{equation}
h_{GBEC}(r)=\left(1-\frac{r^{2}}{R^2_{core}}\right), 
\end{equation}
where the constant $C$ is found when we impose the continuity between the metric functions of DS (GBEC) and the region
of ordinary matter (baryonic matter) described by the Schwarzschild metric. According to the model of Mazur and Motola\cite{16}\cite{17},
we have

\begin{equation}
f(r)\approx\frac{W_{core}}{W}f(R_{core})=C\frac{W_{core}}{W}\left(1-\frac{R^2_{g}}{R^2_{core}}\right)=1-\frac{2Gm}{c^2r},
\end{equation}
where $R_{g}$, $m$, $r$, $R_{core}$ and $W$ are respectively the radius from where the phase transition begins, the total mass of the
star, the variable radius and the radius of the dark core; $W=W(r)=8\pi G r^2 p/c^4$ being a dimensionless variable, where
$W_{core}=8\pi G R_{core}^2 p/c^4$\cite{16}\cite{17}. $G$ is the Newton gravitational constant. However, according to the present model,
we will show that the phase transition occurs approximately at $R_{core}$ (Fig.5). 

The Schwarzchild metric can be written as the DS-metric, so that their metric functions are

\begin{equation}
f_{S}(r)=1-\frac{2Gm}{c^{2}r},~ h_{S}=\frac{1}{\left(1-\frac{2Gm}{c^{2}r}\right)}. 
\end{equation}

The vacuum energy density is

\begin{equation}
\rho=\frac{\Lambda c^2}{8\pi G},
\end{equation}
where $\Lambda$ is the cosmological constant whose vacuum state is represented by the following EOS: 

\begin{equation}
 p=w\rho=-\rho,
\end{equation}
where $w=-1$, $p$ is the pressure and $\rho$ is the vacuum energy density.

On the other hand, the baryonic region with ultra-relativistic matter is described by EOS, namely: 

\begin{equation}
p=w\rho=\rho,
\end{equation}
with $w=+1$.

Here it is important to call attention to the fact that there is a general model for gravastar, i.e., a Dark Energy Star model,
where we have a more general EOS that encompasses several vacuum models, namely:
$\frac{dp}{d\rho}=w(v)=-\omega^{2}=-\frac{v^2}{c^{2}}$\cite{11}\cite{12}. We aim to consider those several vacuum vibration modes inside
GBEC in order to justify better the origin of anisotropy that leads to the increase of the pressure inside GBEC, allowing the 
emergence of a phase transition that finishes to occur at the radius of core from where a Quark Gluon Plasma (QGP) phase begins (Fig.5).

\subsection{TOV of a gravastar and the structure equations}

 By adopting the spherical coordinates, we write a metric with the observables of the star\cite{9}\cite{11}\cite{19}, namely:

\begin{equation}
 ds^{2}=-c^{2}exp\left[-2{\int}^{\infty}_{r}g(r')dr'\right]dt^{2}+\frac{dr^{2}}{\left(1-\frac{2Gm}{c^{2}r}\right)}+r^{2}d\Omega^{2}, 
\end{equation}
where $d\Omega^{2}=d\theta^{2}+\sin^{2}(\theta)d\phi$ is the solid angle. The function $\frac{2Gm}{c^2r}$ is known as compactness.
Such a function measures how relativistic is a star and plays a special role in the structure of a gravastar when we discuss the energy
conditions NEC, DEC, SEC and WEC\cite{13}, which are responsible for determining the closing of a region associated with the formation
of an event horizon.

The compactness is expressed by a ratio of speeds, that is to say we can write $v^2/c^2=\frac{2Gm}{c^2r}$, where $v$ is the speed of
escape from the star. When $v$ is compared to $c$, we should have $v/c<1$ such that light can escape and so the event horizon
is not formed.

We choose the function $g(r)$ to represent the local gravitational acceleration and $m(r)$ is the mass-energy confined into a radius $r$.

The stress-energy tensor is $T^{\mu}_{\nu}=diag[-\rho,p_{r},p_{t},p_{t}]$. The component $00$ of the field\cite{9} is

\begin{equation}
 m(r)c^2=4\pi\int^{r}_{0}\rho(r')r'^{2}dr',
\end{equation}
where $\rho$ is an energy density.

The spatial component $rr$\cite{13} is given by 

\begin{equation}
g(r)=\frac{m(r)c^2+4{\pi}p_{r}(r)r^{3}}{r^{2}\left[1-\frac{2Gm(r)}{c^{2}r}\right]}=
\frac{4{\pi}r}{3}\frac{\rho(r)_{t}+3p_{r}(r)}{\left(1-\frac{2Gm(r)}{c^{2}r}\right)},
\end{equation}
where $\rho(r)=\frac{m(r)c^2}{\frac{4}{3}{\pi}r^{3}}$.

Let us continue to indicate explicitly the dependence of all relevant functions with the radius $r$. Considering the existence of
a transverse pressure $p_{t}$, the more general TOV equation is written as

\begin{equation}
\frac{dp_{r}}{dr}=-(\rho+p_{r})\frac{g}{c^2}+2\frac{(p_{t}-p_{r})}{r}.
\end{equation}

In the isotropic case, we find $p=p_{r}=p_{t}$, however the works of CFV and Debenedicts et al\cite{9}\cite{10}\cite{11}\cite{13} 
demonstrate the relevance of the tangencial (transverse) pressure for a gravastar. So we define the dimensionless parameter that we call 
anisotropy, namely:

\begin{equation}
\Delta=\frac{p_{t}-p_{r}}{\rho}.
\end{equation}

Now we write the TOV equation, which represents the hydrostatic equilibrium of a fluid. It is given in terms of the anisotropy, namely:

\begin{equation}
\frac{dp_{r}}{dr}=-(\rho+p_{r})\frac{g}{c^2}+2\rho\frac{\Delta}{r}.
\end{equation}

In the center of the core of the GBEC ($r=0$), there is no anisotropy ($\Delta=0$). 

To complete the set of the structure equations of a star, we present the continuity equation, which regulates the mass behavior, namely:

\begin{equation}
 \frac{dm}{dr}=\frac{4{\pi}}{c^2}r^{2}\rho.
\end{equation}

\begin{figure}
\begin{center}
\includegraphics[scale=0.80]{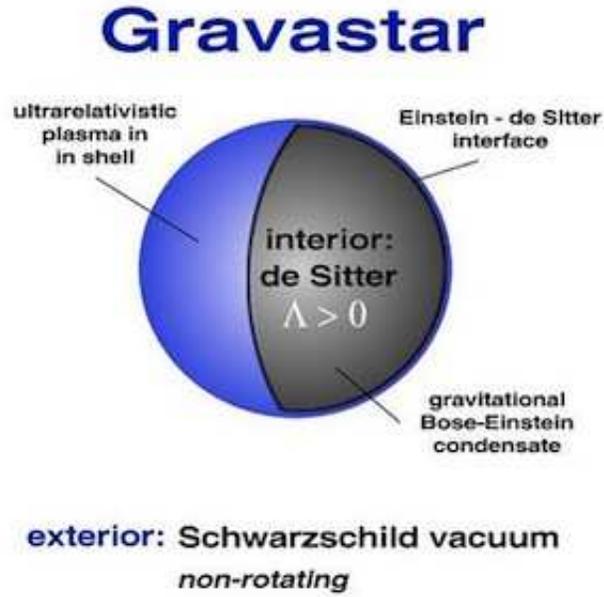}
\end{center}
\caption{In the interior of a gravastar, we find a repulsive core (GBEC) associated with a positive cosmological constant
(anti-gravity)\cite{9}. GBEC is covered with a thin baryonic shell described by the Schwarzschild metric.}
\end{figure}

\subsection{The relationship of DEC and WEC with the anisotropy and compactness}

The study of fluid with negative pressure in stars was regulated in refs.\cite{9}\cite{10}\cite{11}\cite{13} by following the
Buchdahl-Bondi relation. We have the following conditions:

1) The NEC (The Null Energy Condition)

\begin{equation}
 \rho+p_{i}\geq 0.
\end{equation}

2) The DEC (The Dominant Energy Condition)

\begin{equation}
 |{p_{i}}|\leq\rho.
\end{equation}

NEC and DEC are associated with the closing conditions\cite{9}\cite{10}\cite{11}\cite{13} or the appearance of an event horizon.
The imposition of such conditions for the compactness result in a range of values that the compactness should obey in such a way that
the horizon event is not formed. Thus, for the existence of a phase transition with the appearance of a repulsive core (GBEC), the values
of compactness must conform to the following ranges in the respective layers near the Schwazschild radius defined by the value CFV\cite{9}, 
respecting NEC and DEC conditions of Buchdahl-Bondi, namely:

\begin{equation}
\frac{8}{9}<\frac{2Gm}{c^{2}r}<1, ~~ \Delta\geq\frac{1}{4}\frac{\frac{2Gm}{c^{2}r}}{1-\frac{2Gm}{c^{2}r}}>0,
\end{equation}
where $\Delta$ is the magnitude of the anisotropy. The coupling between the compactness and anisotropy characterizes the need to prevent
the formation of the event horizon.

Usually, in the literature about nuclear Astrophysics, the anisotropy is related to the presence of an electromagnetic field, but here
the situation is different since the anisotropic term is introduced so that we can obtain a repulsive effect capable of preventing the
formation of the event horizon. This is why the connection between anisotropy and the compactness is essential, which means that the
anisotropy arises to respect the values above, being in accordance with the compactness that must be $\frac{2Gm}{c^{2}r}<1$.

\section{Mapping between DS-metric that governs GBEC and SSR metric}

In this section we map the geometric structure of SSR\cite{3}\cite{4}\cite{5}\cite{6} in the geometry of spacetime of the GBEC
inside a gravastar. So we will rewrite DS-metric (Eqs.(7), (8), (9) and refs.\cite{16}\cite{17}). In view of this, here let us
consider that there are effective constants like $c'$\cite{18}\cite{25}, $V ',\Lambda'$ and $H'$ for representing respectively an
effective speed of light and a minimum speed in the GBEC medium, and also an effective cosmological constant connected to an effective
Hubble constant associated with the spacetime of a certain GBEC in the core of a gravastar. Thus, we write 

\begin{equation}
 ds^{2}=-f(r)c'^{2}dt^{2}+\frac{dr^{2}}{h(r)}+r^2d\Omega^{2},
\end{equation}
where the metric functions are
\begin{equation}
f_{DS}(r)=C\left(1-\frac{r^2}{R^{2}_{core}}\right),~ h_{DS}(r)=\left(1-\frac{r^2}{R^2_{core}}\right).
\end{equation}

We also rewrite the spacetime metric of SSR\cite{3}\cite{4}\cite{5}\cite{6} in the following way:

\begin{equation}
ds^{2}=-\frac{c'^{2}dt^{2}}{\left(1-\frac{V'^{2}}{v^{2}}\right)}+\frac{dr^{2}}{\left(1-\frac{V'^{2}}{v^{2}}\right)}+
r^{2}d\Omega^{2}\frac{1}{\left(1-\frac{V'^{2}}{v^{2}}\right)}.
\end{equation}

\begin{figure}
\begin{center}
\includegraphics[scale=0.80]{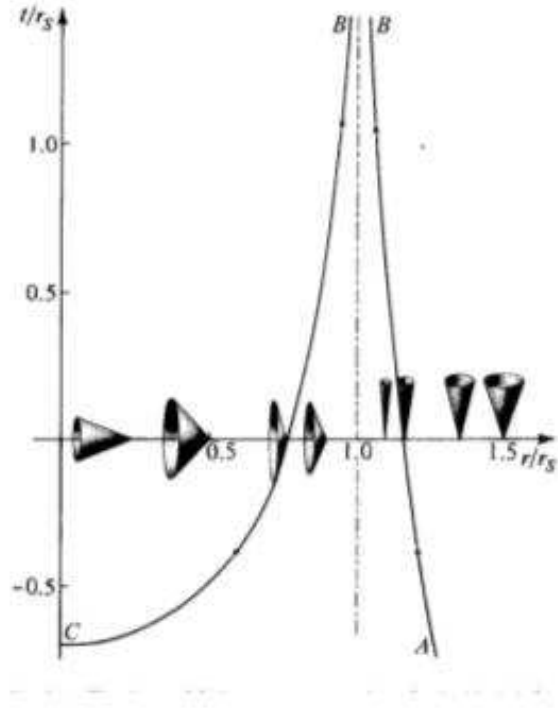}
\end{center}
\caption{This figure shows a usual causal structure of black hole (BH) with event horizon\cite{18}. This structure will be modified with the
introduction of the
causal structure of SSR\cite{3}\cite{19}. We see that the speed of light changes close to the event horizon when the gravitational field 
is extremely high. Thus, we expect that the speed of light, the minimum speed and the cosmological constant also acquire
specific values in a gravitational collapse. The formation of a gravastar due to a phase transition\cite{9} that leads to the emergence
of GBEC is connected to a new structure of spacetime having a positive cosmological constant (a highly repulsive core). This allows us to 
believe that such fundamental greatness like $c$, $V$ and $\Lambda$ could really change in the GBEC of a gravastar, although we will
explore deeper such a change elsewhere. In the last section, we will show how $c'$ and $V'$ vary just in the region of phase transition.}
\end{figure}

\subsection{Component $rr$ of the metric}

By equating the two metric functions associated with $rr$ component of the metric (Eq.(6) and Eq.(10)), we find

\begin{equation}
\frac{1}{\left(1-\frac{r^2}{R^2_{core}}\right)}=\frac{1}{\left(1-\frac{V'^{2}}{v^{2}}\right)},
\end{equation}

from where, we get

\begin{equation}
\frac{V'^{2}}{v^{2}}=\left[\frac{r}{R_{core}}\right]^{2}.
\end{equation}

And knowing that $\rho=\frac{{\Lambda'}c'^{2}}{8\pi G}$, with $\Lambda'={3H'^{2}}$, we write

\begin{equation}
\frac{V'^{2}}{v^{2}}=\frac{8G\rho{\pi}r^{2}}{3c'^{4}}. 
\end{equation}

The above equation shows the connection between the minimum speed and the dark mass density of a gravastar. Thus, we are able to realize
that, since the minimum speed $V'$ inside the GBEC depends on the density $\rho$ of the star, indeed such a minimum speed $V'$ is not 
necessarily the fundamental minimum speed $V$ obtained for the universe\cite{3}. As the density $\rho$ of the GBEC is normally much higher
than the fundamental vacuum energy density of the universe, of course we expect that $V'>V$ so that each star has its value of $V'$, 
whereas $V$ is a universal constant\cite{3}. So, as it is already known that a too strong gravity generated inside BH\cite{18} can lead to
a variation of the speed of light (Fig.4), it would be also natural to expect that there is an effective minimum speed $V'$ inside a 
system on gravitational collapse, preventing the appearence of the singularity of horizon event and thus justifying physically the 
emergence of GBEC. The effective minimum speed $V'$ and the speed of light $c`$\cite{18} (Fig.4) given in the region of phase 
transition of a gravastar will be obtained in section 5. 

Rewriting the term of minimum spped ($V'^{2}/v^{2}$) in function of the effective cosmological constant given in the core of a gravastar,
we find 

\begin{equation}
\frac{V'^{2}}{v^{2}}=\frac{{\Lambda'}r^{2}}{3c'^{2}}. 
\end{equation}

This result provides interesting consequences, since there is an upper limit for $\alpha=\frac{V'^{2}}{v^{2}}$, i.e., we always find
$\alpha<1$ in view of the fact that $V'$ corresponds to the unattainable minimum limit of speed in the region of GBEC, and thus we can
also obtain a limit for the radius of GBEC of a gravastar, which represents the radius of core $R_{core}$. Beyond this, we should
consider the radial function $v=v(r)$ so that we can study the propagation of signals in the interior of the GBEC, where there should be
many kind of vibrating vacuums, also including the special case of the cosmological constant ($p=-\rho$ with $w=-1$).

By manipulating the above equation, we can find the following dimensionless constant for a gravastar (GBEC), i.e., $\xi'=\frac{V'}{c'}$,
namely:

\begin{equation}
\xi'^{2}=\frac{{\Lambda'}v^{2}r^{2}}{3c'^{4}}.
\end{equation}

The value of compactness connected to the cosmological constant is in agreement with the value that was found in previous works about
gravitational collapses\cite{7}\cite{8}.

\subsection{The continuity of the metric in the region of phase transition}

The metrics of the two regions of a gravastar must obey the condition of continuity in the region of phase transition\cite{16}\cite{17}.
Taking into account such a condition, let us impose this condition to both metrics of SSR and Schwarzschild (Eq.(6) and Eq.(18)), by
begining with the component $rr$. The Schwarzschild metric is

\begin{equation}
ds^{2}=-c'^{2}exp[-\frac{2}{c'^2}\int^{\infty}_{r}g(r')dr']dt^{2}+\frac{dr^{2}}{1-\frac{2Gm}{c'^{2}r}}+r^{2}d\Omega^{2}.
\end{equation}

The condition of continuity applied to the component $rr$ of both metrics on about the radius of core $r{\approx}R_{core}$, where the
phase transition occurs, leads to

\begin{equation}
\frac{1}{\left(1-\frac{V'^{2}}{v^{2}}(r_{\approx} R_{core})\right)}=\frac{1}{\left(1-\frac{2Gm}{R_{core}c'^{2}}\right)}.
\end{equation}

So we find that the ratio of speeds is directly connected to the compactness, as follows:

\begin{equation}
\left(\frac{V'}{v(r{\approx}R_{core})}\right)^{2}=\frac{2Gm}{R_{core}c'^{2}},
\end{equation}
where we should stress that the mass of such a gravastar ($m$) is considered to be approximately equal to its dark mass (core mass), since
its baryonic shell is too thin so that we are making the following approximation: $m_{core}\approx m_{star}=m$. 

This result is expected since it is already known that the own compactness could be written as a ratio of speeds. Thus we notice that
$v^{2}_{esp}={\frac{2Gm}{r}}$ gives the escape velocity of a star. Therefore we realize that the compactness could be understood as being
the ratio $(v^{2}_{esp}/c^{2})$. So, with the imposition of the Buchdal-Bondi relation\cite{7}\cite{8}, we write

\begin{equation}
\frac{8}{9}<\frac{V'^{2}}{v^{2}(r{\approx}R_{core})}<1.
\end{equation}

We finally obtain the upper limit for the ratio of speeds. Although such a limit was already predicted before, here we have
another limit, which also indicates when the phase transition begins to occur. This condition imposed to the compactness is
compatible with the impossibility of formation of event horizon. Thus we have a strong evidence that the causal structure of SSR prevents
the formation of event horizon, even because the minimum speed is unattainable such that we must have $V'/v<1$, which is in agreement
with the Buchdahl-Bondi relation, as we get $v(r)>V'$. In this case, we interpret $v$ as being the input speed into the GBEC, that is
the opposite of the idea of escape speed from gravity, since we are considering that the core of a gravastar represents a repulsive region
connected to an anti-gravity, so that now we need an input speed $v(r)$ to reach the interior of the repulsive core (GBEC) for a certain 
radial coordinate $r<R_{core}$.

Let us now apply the continuity condition for the temporal component $00$. As we are on the surface of the dark core governed by GBEC, so
we are in the limit of validity of the causal structure of SSR. Imposing the continuity condition to the components $00$ of both metrics
of SSR (Eq.(6)) and Schwarzschid (Eq.(18)), we obtain

\begin{equation}
\frac{1}{\left(1-\frac{V'^{2}}{v^{2}(R_{core})}\right)}=-exp\left[-\frac{2}{c'^2}\int^{\infty}_{r\approx R_{core}}g(r')dr'\right], 
\end{equation}
where we consider $r{\approx}R_{core}$ (on the shell of the dark core).

and so we find

\begin{equation}
-\frac{2}{c'^{2}}\int^{\infty}_{r{\approx}R_{core}}g(r)dr=ln\left|-\frac{1}{\left(1-\frac{V'^{2}}{v^{2}(R_{core})}\right)}\right|
\end{equation}

or yet

\begin{equation}
 =-\ln\left(1-\frac{V'^{2}}{v^{2}(r{\approx}R_{core})}\right). 
\end{equation}

By deriving in $r{\approx}R_{core}$ for both of sides of the above equation, we obtain

\begin{equation}
g(\infty)-g(R_{core})=\frac{c'^2}{\left(1-\frac{V'^2}{v^2}\right)}\frac{V'^{2}}{v^{3}}\frac{dv(r)}{dr}|_{r{\approx}R_{core}}.
\end{equation}

As $g(r)$ represents a gravitational acceleration in the region of Schwarschild, we have $g(\infty)=0$ so that we find

\begin{equation}
g(r)=-\frac{c'^2}{\left(1-\frac{V'^2}{v^2}\right)}\frac{V'^{2}}{v^{3}}\frac{dv}{dr}.
\end{equation}

Knowing that $v=\frac{V'R_{core}}{r}$, we find $\frac{dv}{dr}=-\frac{V'R_{core}}{r^{2}}<0$.

In the same way we can write a condition for the ratio of speeds that prevents the event horizon, we can also write a ratio of
possible radius of core in such a way to respect the causality condition by preventing the closing or formation of the event horizon,
and thus forming a phase transition as suggested by Chapline\cite{9}\cite{10}\cite{12} (see Fig.5); however this question is still to be
better explained. We have

\begin{equation}
\frac{8}{9}<\left[\frac{R_{g}}{R_{max}}\right]^{2}<1.
\end{equation}

We believe that $R_{core}$ (GBEC radius) vary by means of some intriguing way in the interval $r_{g}<R_{core}<R_{max}$ (Fig.5). Since 
such a phase transition is preventing the formation of the coordinate singularity, the radius of GBEC must assume a value
$R_{core}<c'\sqrt{\frac{3}{\Lambda'}}$.

 It is interesting to notice that all the results we have obtained show explicitly the coupling of a certain cosmological parameter with
the greatnesses associated with SSR\cite{3}.

\subsection{The Equation of State (EOS) and the energy conditions}

  A gravastar has an EOS\cite{9}\cite{10}\cite{16}\cite{17}. By considering the more generic case of vacuum, let us think about many 
possible effects of repulsive pressure of a Gravastar/Dark Energy Star, being represented by the following EOS for GBEC:

\begin{equation}
 p=-\frac{v^{2}(r)}{c'^{2}}\rho,
\end{equation}
where $v=v(r)$ was interpreted as an input speed into the GBEC with a certain radial coordinate $r$. We should note that, if the input 
speed is $v\rightarrow V$, we would be in limit of validity of SSR, which corresponds to the radius $r=R_{core}$ of the shell of the dark 
core. And, if the input speed is $v\rightarrow c$, we would be very close to the center of the dark core ($r\rightarrow 0$), which is 
governed by EOS of the cosmological constant, where we have $p=-\rho$. In short, SSR leads us to deal with the core of a gravastar
as being made by a set of infinite spherical shells connected to a continuum spectrum of modes of vacuum, instead of simply only one mode
of vacuum connected to the cosmological constant as described in the simple model in ref.\cite{9} (Fig.3).

The energy conditions DEC and NEC being revisited (Eq.(25) and Eq.(26)) are given, as follows:

\begin{equation}
 \rho+p_{i}\geq 0,
\end{equation}

and

\begin{equation}
 \rho\geq|{p}|.
\end{equation}

By using the density $\rho$ calculated before (Eq.(46)), we find the following EOS: 

\begin{equation}
\rho=\frac{3V'^{2}c'^4}{8\pi Gr^{2}v^{2}(r)}.
\end{equation}

Thus, by using EOS above, we rewrite the pressure in terms of the ratio of speeds, where we consider the causal structure of SSR
being directly related to the question about the formation of event horizon. So, if the energy conditions DEC and NEC are obeyed, this
ensures that the event horizon does not appear for any value of radius, especially for $r{\sim}R_{core}$. Then we write

\begin{equation}
p=-\frac{3V'^{2}c'^2}{8\pi Gr^{2}}
\end{equation}

By substituting the above result (Eq.(50)) into the energy condition, we find

\begin{equation}
\frac{\Lambda'c'^{2}}{8\pi G}-\frac{3V'^{2}c'^2}{8\pi Gr^{2}}\geq 0.
\end{equation}

So we verify that

\begin{equation}
\Lambda'\geq 3\left(\frac{V'}{r}\right)^2,
\end{equation}
where there must be a minimum radius $r=r_{min}>0$ in order to avoid the central singularity, i.e., in order to prevent 
that $\Lambda'\rightarrow\infty$. This question will be discussed elsewhere.

It is important to stress that there is a strong connection between an effective cosmological constant $\Lambda'$ and an effective
mimimum speed $V'$ given in spacetime of the region of GBEC. We also realize that the minimum speed $V'$ is directly proportional
to the repulsive pressure. So, if $V'=0$, this would imply in a null pressure (a null anti-gravity), which would lead to the absence 
of the dark core and thus the appearence of event horizon of a BH.

Now verifying the other condition, i.e., $\frac{\Lambda c'^{2}}{8\pi G}\geq\frac{3V'^{2}c'^2}{8\pi Gr^{2}}$, we recover again the
same result above. This strengthens our conclusion that there is no event horizon, although we do not still have a demonstration
for the inferior limit that appears in the Buchahl-Bondi relation. In sum, we have shown that SSR obeys the energy condition for
the upper limit $1$ in the Buchahl-Bondi relation, but we need to develop a new energy condition based on SSR in order to investigate
the lower limit $8/9$ that appears in the Buchahl-Bondi relation. This issue will be explored elsewhere.

If we consider that the effective minimum speed $V'$ is a constant inside the core of a gravastar, we find the pressure on the surface 
of the dark core. So we have

\begin{equation}
V'=R_{core}\sqrt{\frac{-8\pi G p(R_{core})}{3}},
\end{equation}
where we should remember that $p<0$. Beyond this, we see a direct dependence between $V'$ and the radius of the dark core ($R_{core}$),
so that, if $V'=0$, the dark core would vanish ($R_{core}=0$), and therefore all this new physics based on SSR vanishes. This
strengthens our comprehension that the quantum spacetime of SSR is needed to give consistency to the formation of a phase trasition that
emerges in order to replace the event horizon by GBEC during the gravitational collapse.

\subsection{TOV and structure equations in terms of ratio of speeds}

\begin{figure}
\begin{center}
\includegraphics[scale=0.80]{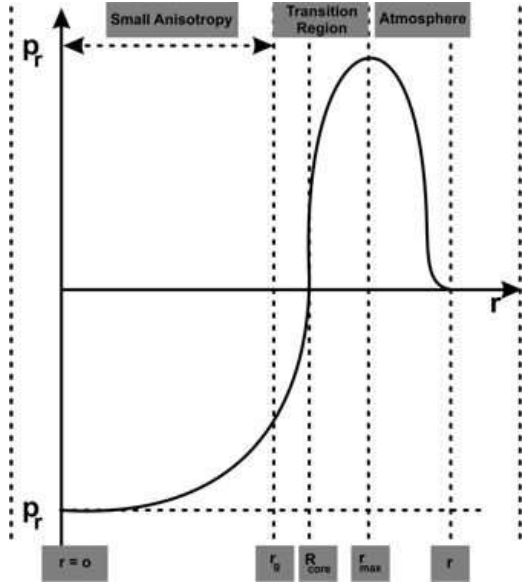}
\end{center}
\caption{Graph of the radial pressure versus radius, showing the region inside which there is a radius of phase transition, i.e., such 
a radius is in somewhere in the interval $r_{g}<r<r_{max}$ to be determined. The dominant anisotropy occurs for
$p(R_{core})\approx 0$\cite{9}\cite{10}, which is close to the region of baryonic matter. Here we intend to show that such a 
transition occurs at a coexistence radius $R_{coexistence}\approx R_{core}$ where the radial pressure $p$ vanishes. In the last section, 
we will show that such a radius is slightly larger than the Schwarzschild radius $R_{S}$, i.e., $R_{S}<R_{core}\approx R_{coexistence}$,
preventing the divergence of the metric. In the interval $R_{core}<r<r_{max}$, the pressure decreases from its maximal value at $r_{max}$
to zero at $R_{core}$ (transition region) due to the fact that gravity is reduced abruptally in this interval, where the baryonic matter
is crushed into a Quark Gluon Plasma (QGP). We believe that the origin of such a rapid decrease of gravity comes from the vacuum 
anisotropy that already begins to govern the collapse for $r<r_{max}$ until reaching the phase transition (gravity/anti-gravity)
at $r\approx R_{core}$, where the anisotropy reaches its maximum value. Thus the anisotropy could be responsible for the emergence of 
different radius like $r_{max}>R_{coexistence}(\approx R_{core})>R_S>r_g$, instead of simply $R_S$ for the classical collapse, where 
we would have $r_{max}=R_S$ with a divergent pressure (no phase transition). A more accurate discrimination of those radius will be 
investigated elsewhere.}
\end{figure}

Let us begin by rewritting a TOV equation (Eq.(23)), which corresponds to the hydrostatic stability of the star. We first consider the
absence of anisotropy ($\Delta=0$), namely:

\begin{equation}
 \frac{dp}{dr}=-(\rho+p)\frac{g}{c'^2}.
\end{equation}

We have already found $p=p(r)$ (Eq.(50)), such that $dp(r)/dr$ is

\begin{equation}
\frac{dp}{dr}=\frac{3V'^{2}c'^2}{4\pi Gr^{3}}.
\end{equation}

We see that $dp/dr>0$, which implies that there should be a repulsive effect. Such a condition is expected for this kind of fluid.

Taking into account EOS of a gravastar, the function $g(r)$ and also the density with respect to $V'$, we find

\begin{equation}
\frac{dp}{dr}=\left(1+\frac{c'^{2}}{v^{2}(r)}\right)\frac{3V'^{4}c'^2}{8\pi G r^{2}}
\frac{1}{\left(1-\frac{V'^2}{v^{2}(r)}\right)}\frac{1}{v^{3}(r)}\frac{dv(r)}{dr}.
\end{equation}

The mapping between DS and SSR metrics (Eq.(33)) is informing us that $\frac{dv(r)}{dr}<0$. In order to get a positive gradient
of pressure in accordance with the result already found before, from the left side of the equation, we need to compatibilize the terms
with negative signal that have origin on the gravitational potential (Eq.(44)), in such a way that we find $\frac{dp}{dr}>0$.

We can still calculate the gradient of the input speed by using Eq.(56) and by comparing it with Eq.(55). According to SSR,
this is a fundamental greatness since SSR connects the uncertainty principle to the concept of non-localization, which has origin 
in the idea of reciprocal velocity that has been studied in a previous paper\cite{5}. Such a gradient $dv(r)/dr$ represents a frequency, 
as follows:

\begin{equation}
\left[\frac{dv(r)}{dr}\right]_{iso}=\frac{1}{r}\frac{\left(1-\frac{V'^{2}}{v^{2}(r)}\right)}
{\left(1+\frac{c'^2}{v^2}\right)}\frac{v^{3}(r)}{V'^{2}}.
\end{equation}

At a first sight, we would have a high level of vibration of GBEC (a high frequency) close to the input speed of light, which means close
to the center of the core, i.e.:

\begin{equation}
 \lim_{v{\rightarrow}c'}\left[\frac{dv}{dr}\right]_{iso}=\frac{(1-\xi'^{2})}{2r}\frac{c'}{\xi'^{2}}.
\end{equation}

In the limit of the input minimum speed $V'$ or close to the surface of GBEB, at a first sight, we would find

\begin{equation}
\lim_{v{\rightarrow}V'}\left[\frac{dv}{dr}\right]_{iso}\rightarrow 0.
\end{equation}

We can interpret the gradient $\frac{dv}{dr}$ as being the vacuum frequency for a certain radial coordinate inside the core (GBEC) of
a gravastar. 

For a speed close to the speed of light, a proof particle is close to the center of GBEC and it vibrates very fast. Therefore this leads 
to the increase of the pressure. On the other hand, close to the limit of the minimum speed (core radius), the pressure tends to 
remain constant; however, as the ultra-referential $S_V$ is inaccessible, the pressure is never constant. We interpret that such vibrations of GBEC
are responsible for variations of the pressure so that the formation of the event horizon does not occur. However we need a pressure that
has a rapid increase in the region of phase transition, being necessary that $[\frac{dp}{dr}]_{r\approx R_{core}}>>0$. This compels us to 
add an anisotropic term and then also associate it with the structure of SSR. Thus, this anisotropic term materializes the emergence of 
the phase transition, since it is the parameterizer of the change of types of vacuum. We will study this subject in the next section.

\subsection{The anisotropic TOV}

 By introducing the anisotropic term in the TOV equation (Eq.(54)) in order to prevent a quasi-null gradient (Eq.(59)) close
to the surface of the dark core, then, at a more accurate sight, we have 
 
\begin{equation}
\frac{2\rho_{eff}}{r}\Delta=\frac{dp_{r}}{dr}-\left(1+\frac{c'^{2}}{v^{2}(r)}\right)
\frac{3V'^{4}c'^4}{8\pi Gr^{2}}\frac{1}{\left(1-\frac{V'^2}{v^{2}(r)}\right)}\frac{1}{v^{3}(r)}\frac{dv(r)}{dr}.
\end{equation}

The anisotropic term needs to obey some conditions introduced by CFV\cite{9}\cite{10} in order to assure the continuity of the metric
that has been anounced in Eq.(27). So we rewrite 

\begin{equation}
\Delta\geq\frac{1}{4}\frac{\frac{(V')^2}{v^2}}{\left(1-\frac{(V')^2}{v^2}\right)}.
\end{equation}

In order to study anisotropy of a super massive system like a gravastar, we need to reintroduce another structure equation (Eq.(19)),
i.e., the mass continuity since mass represents a gravitational charge so that its distribution can be necessary for the phase 
transition we are looking for. We find

\begin{equation}
\frac{dm}{dr}=\frac{4\pi}{c'^{2}}r^{2}\rho.
\end{equation}

By using the EOS, i.e.,$p=-\frac{v^{2}(r)}{c'^{2}}\rho$, so we have 

\begin{equation}
\frac{dm}{dr}=-4\pi r^{2}\frac{p(r)}{v^{2}(r)}.
\end{equation}

By deriving this gradient of mass above (Eq.(63)), we find 

\begin{equation}
\frac{d^{2}m}{dr^{2}}=-4\pi\left(2r\frac{p(r)}{v^{2}(r)}\right)+
\left[\frac{v^{2}\frac{dp(r)}{dr}-2v(r)p(r)\frac{dv(r)}{dr}}{v^{4}(r)}\right]r^{2}.
\end{equation}

After some mapulations else, we find 

\begin{equation}
\frac{v^{2}(r)}{4\pi r^{2}}\frac{d^{2}m}{dr^2}-\frac{2p(r)}{r}=\frac{dp(r)}{dr}-\frac{p(r)}{v(r)}\frac{dv(r)}{dr}.
\end{equation}

So we have a profile for anisotropy, i.e., $\frac{2\rho_{eff}\Delta}{r}=\frac{v^{2}(r)}{4\pi r^{2}}\frac{d^{2}m}{dr^2}
-\frac{2p(r)}{r}=\frac{2(p_t-p_r)}{r}$, which meets the requirements of continuity of the metric (Eq.(61)). By considering these
 requirements, where we have $\Delta(r=0)=0$, we must redefine the anisotropy factor as a dimensionless term\cite{9}\cite{10}. We have

\begin{equation}
\Delta=\frac{p_{t}-p_{r}}{\rho}, 
\end{equation}
where $p_{t} $ and $p_{r}$ are respectively the tangential and radial pressures. Our EOS is given in terms of the radial pressure, i.e.,
$p_{r}=-\frac{v^{2}(r)}{c'^{2}}\rho$.

By inspecting Eq.(65) above, we conclude that the tangential pressure is given by

\begin{equation}
 p_{t}(r)=\frac{v^{2}}{8\pi r}\frac{d^{2}m(r)}{dr^{2}}.
\end{equation}

As we want to associate the tangential pressure with the change in mass ($dm(r)^2/dr^2$), here we link the second
derivative $dm(r)^2/dr^2$ with the anisotropic term. This coupling ensures that close to the surface of the dark core, 
we have $[\frac{dp(r)}{dr}]_{r\approx R_{core}}>>0$. In view of this engagement, we have to impose conditions on the second derivative of 
mass. Here SSR can also bring an important contribution since Nassif\cite{4} brought the concept of dressed mass and anisotropy of
the mass, which are generated just by the presence of the ultra-referential (vacuum energy) that dresses strongly the mass of a particle close 
to the minimum speed, where a mass of vacuum (dressed mass) close to the minimum speed was obtained as being 
$m=\frac{m_{0}}{\sqrt{1-\frac{V'^{2}}{v^{2}(r)}}}$\cite{4}. The second derivative of such effective (dressed) mass when $v$ tends to the 
minimum speed is given by

\begin{equation}
\frac{d^{2}m_{dress.}(r)}{dr^{2}}=m_{0}\left(\frac{dv}{dr}\right)^{2}_{eff}
\left[\frac{3V'^{4}}{2v^{6}}\frac{1}{\left(1-\frac{V'^{2}}{v^{2}(r)}\right)^{5/2}}\right].
\end{equation}

The approximation given in Eq.(68) obeys the requisition established by CFV\cite{9}\cite{10}, which was announced in Eq.(61).

Eq.(68) has origin in the Lorentz symmetry breaking due to an invariant minimum speed, which justifies the emergence of a strong 
anisotropy in the region of phase transition. 

Based on Eq.(68), we can say that $m_{0}\left(\frac{dv}{dr}\right)^{2}_{eff}=r\rho_{eff}$, where $\left(\frac{dv}{dr}\right)^{2}_{eff}$
is a gradient associated only with the anisotropic part, so being an effective gradient. Finally we write our approximation for the
anisotropy, namely:

\begin{equation}
 \Delta(v\approx V')=\frac{3}{4}\frac{\left[\frac{V'}{v(r)}\right]^{4}}{\left(1-\frac{V'^{2}}{v^{2}(r)}\right)^{5/2}}.
\end{equation}

In Eq.(69), we note that the exponents are different from those established by CFV\cite{9}\cite{10}. Here, SSR is able to show a 
quantitative difference with regard to the ansatz proposed by CFV\cite{9}\cite{10}, leading to a more accurate result for anisotropy. 

We reiterate that the fact that the tangential pressure is proportional to the second derivative of the mass brings out precisely the
fact that we are dealing with a ultra-massive object (of type BH), so that small asymmetries in the mass distribution can result
in changes of the distribution of gravitational fields. In fact, GBEC consists of a type of superfluid associated with vacuum\cite{3}\cite{12}.
These asymmetries introduced by tangential pressure $p_t$ are actually asymmetries in vacuum due to various types of vacuum
that constitute the core\cite{3}\cite{12}, whose surface (for $r= R_c$) represents a coexistence region of phase, which is the boundary
between the regions of pure vacuum described by SSR and the region of attractive gravity, being usually filled by baryonic matter
that we do not explore in this paper.

Let us now write the entire TOV with the term of anisotropy by substituting Eq.(65), Eq.(68) and EOS into Eq.(60) and thus obtaining 

\begin{equation}
\frac{dp_{r}(r)}{dr}=\left(1+\frac{c'^{2}}{v^{2}(r)}\right)\frac{3V'^{4}c'^4}{8\pi Gr^{2}}
\frac{1}{\left(1-\frac{V'^2}{v^{2}(r)}\right)}\frac{1}{v^{3}(r)}\frac{dv(r)}{dr}
+\frac{v^{2}}{2\pi r^2}\frac{d^{2}m(r)}{dr^{2}}-2\frac{v^{2}(r)\rho}{rc'^{2}}.
\end{equation}

In Eq.(70), when we place explicitly the anisotropic term, we find 

\begin{equation}
\frac{dp_r(r)}{dr}-\left(1+\frac{c'^2}{v^2(r)}\right)
\frac{3V'^4 c'^4}{8\pi Gr^2}\frac{1}{\left(1-\frac{V'^2}{v^2(r)}\right)}
\frac{1}{v^3(r)}\frac{dv(r)}{dr}=\frac{v^2}{2\pi r^2}\frac{d^2 m(r)}{dr^{2}}-2\frac{v^2(r)\rho_{eff}}{c'^2r}.
\end{equation}

Writing the second derivative of mass in function of $\rho_{eff}$, we obtain

\begin{equation}
\frac{d^{2}m(r)}{dr^{2}}=
\left[\frac{3V'^{4}}{4v^{4}}\frac{1}{\left(1-\frac{V'^{2}}{v^{2}(r)}\right)^{5/2}}\right]\frac{r\rho_{eff}}{c'^{2}}.
\end{equation}

Substituting Eq.(55) and Eq.(72) into the Eq.(71), we finally get the TOV-SSR equation for the gradient of the speed of propagation, 
namely: 

\begin{equation}
\left[\frac{dv(r)}{dr}\right]_{aniso}=-\frac{2\left(1-\frac{V'^2}{v^{2}(r)}\right)v^{3}}{rV'^{2}\left(1+\frac{c'^{2}}{v^{2}}\right)}
-\frac{8r\rho_{eff}Gv^{5}}{3V'^4c'^4}\frac{\left(1-\frac{V'^2}{v^2}\right)}{\left(1+\frac{c'^2}{v^2}\right)}\left(\frac{3V'^{4}}{4v^{4}}
\frac{1}{\left(1-\frac{V'^{2}}{v^{2}(r)}\right)^{5/2}}\right)-\rho_{eff}\frac{\left(1-\frac{V'^2}{v^2}\right)}
{\left(1+\frac{c'^2}{v^2}\right)}\frac{32{\pi}rGv^5}{3V'^{4}c'^6}. 
\end{equation}

We notice that the anisotropic part and the isotropic part are acting as cutoff of each other. This is an important result of 
this work. We still think that $r=\frac{V'R_{core}}{v}$, where we have $v=v(r)$ so that $v$ never vanishes due to the minimum speed, and
$v$ never goes to infinity as $v$ is limited by $c'$ as the upper limit. Thus, as we have two cut-offs such as $V'$ and $c'$, the two 
singularities (both horizon and central singularities) are excluded, since $0<[dv/dr]_{aniso}<\infty$. 

Manipulating Eq.(73), in order to put the isotropic and anisotropic terms in an explicit way, we find 

\begin{equation}
\left[\frac{dv(r)}{dr}\right]_{aniso}=-\frac{(1-\frac{V'^2}{v^{2}(r)})4v^{3}}{rV'^{2}
(1-\frac{c'^{2}}{v^{2}})}[1-\frac{8r^{2}\rho_{eff}Gv^{2}}{3V'^{2}c'^4}[(\frac{3V'^{4}}{4v^{4}}
\frac{1}{(1-\frac{V'^{2}}{v^{2}(r)})^{\frac{5}{2}}})+\frac{2{\pi}}{c'^{2}}]]. 
\end{equation}

Now, by placing the anisotropic factor in Eq.(74) in the explicit form, we obtain

\begin{equation}
[\frac{dv(r)}{dr}]_{aniso}=-[\frac{dv}{dr}]_{iso}[[1-\frac{8r^{2}\rho_{eff}Gv^{2}}{3V'^{2}c'^4}
[(\frac{3V'^{4}}{4v^{4}}\frac{1}{(1-\frac{V'^{2}}{v^{2}(r)})^{\frac{5}{2}}})+\frac{2{\pi}}{c'^{2}}]]. 
\end{equation}

In Eq.(75), there emerges the effective density $\rho_{eff}$, arising from the anisotropy factor, which is associated with the mass in
the coexistence region between the two phases, i.e., $m(r)=m(R_c)\approx m$. Thus, within this more accurate sight, the ``frequency of
vacuum'' for $r=R_c$ ($[dv/dr]_{aniso,r\approx R_c}$) does not vanish in the region of phase transition as would occur if we just consider 
the isotropy, i.e., $[dv/dr]_{iso}$. In sum, from the point of view of SSR, the frequency of vacuum given by the anisotropic factor
($[dv/dr]_{aniso}$) within such a more accurate sight is now capable of preventing the formation of the event horizon at the 
Schwarzschild radius $R_S$, which is now replaced by the core radius $R_c$. 

Later we will show that the coexistence radius ($R_{core}=R_c$) between the two phases must be slightly larger than the Schwarzschild
radius in order to prevent the singularity exactly at $R_S$ and therefore allowing the emergence of such a phase transition from gravity
(baryonic phase) to anti-gravity (GBEC inside the core) at $R_c$. So we must have $R_c>R_S$ to be investigated later. 

Actually, we should conclude that, in this non-classical model of gravitational collapse (gravastar), the idea of Schwarzschild radius
$R_S$ of a star does not make sense since $R_S$ is the result of a singularity generated by a classical gravitational collapse, where 
there is no phase transition.

\subsection{Relationship between gradient of speed and quantum structure of GBEC}

In the development of this article, in three different opportunities, we find the quantity $\frac{dv(r)}{dr}$ having the dimension of
frequency, which was obtained more explicitly in the calculation of the isotropic TOV. We find that this quantity is related to a very
similar structure to the wavelength calculated in SSR\cite{5}, where it was shown that the relativistic effects close to the 
minimum speed lead to a stretching of the de-Broglie wavelength $\lambda_{SSR}(r)$, by reducing the frequency. This is the quantum
analog of the dilation effect of space and time contraction in SSR, which leads to the emergence of the uncertainty principle\cite{5}. 
So we have\cite{5}:

\begin{equation}
\lambda_{SSR}(r)=\frac{h}{m_{0}v(r)}\frac{\sqrt{1-\frac{v^{2}(r)}{c'^{2}}}}{\sqrt{1-\frac{V'^{2}}{v^{2}(r)}}}. 
\end{equation}

In Eq.(76), by isolating the factor $\theta$ of SSR, we find 

\begin{equation}
\frac{1}{1-\frac{V'^{2}}{v^{2}(r)}}=\lambda^{2}(r)\frac{v^{2}(r)m^{2}_{0}}{h^{2}(1-\frac{v^{2}(r)}{c'^2})}.
\end{equation}

Substituting Eq.(77) into Eq.(57), we obtain: 

\begin{equation}
[\frac{dv(r)}{dr}]_{iso}=-\frac{vh^{2}}{rm^{2}_{0}c'^{2}V'^{2}\lambda^{2}_{isoSSR}}
\frac{1-\frac{v^{2}(r)}{c'^{2}}}{1+\frac{c'^{2}}{v^{2}(r)}}. 
\end{equation}

Here $m_{0}$ is the mass of a proof particle that oscillates in the GBEC with frequency $[\frac{dv(r)}{dr}]_{iso}$ and having 
wavelength $\lambda_{iso}(r)$.  

Analyzing Eq.(78), we realize that, when $v{\rightarrow}V'$, we have $\lambda_{iso}\rightarrow\infty$, which means there is a complete 
delocalization of the particle, leading to $[\frac{dv}{dr}]_{iso}\rightarrow 0$, i.e .:

\begin{equation}
[\frac{dv(r)}{dr}]_{iso}|_{v(r){\approx}V'}=-\frac{{\xi'}h^{2}}{c'V'^{2}\lambda(r{\approx}R_{core})}
\frac{1-\xi'^{2}}{1+\frac{1}{\xi'^{2}}}.
\end{equation}

Eq.(79) represents an approximation for a proof particle that is very close to the surface of the core (coexistence region between 
attractive and repulsive phases of gravitation). This corresponds to be close to the ultra-referential $S_{V'}$, $V'$ being unattainable
and therefore the energy cannot be transmitted since the surface of the dark cone inside the star ($v=V`$) is an interval of space-like
(Fig.1)\cite{3}\cite{4}, where the module of $4$-vector ($ds$) tends to infinity\cite{3}, leading to bizarre quantum effects 
like non-locality.

Knowing that $\rho_{eff}=\frac{{\Lambda'_{eff}}c'^{2}}{8{\pi}G}$, and by substituting Eq.(77) into Eq.(75), we write the anisotropic
gradient of speed, namely: 

\begin{equation}
[\frac{dv}{dr}]_{aniso}=-[\frac{dv}{dr}]_{iso}[1-\frac{8r^{2}\rho_{eff}Gv^{2}}{3V'^{2}c'^{4}}
[\frac{3V'^{4}}{2v^{4}(r)}\frac{\lambda^{5}m^{5}_{0}v^{5}(r)}{h^{5}(1-\frac{v^{2}(r)}{c'^{2}})^{\frac{5}{2}}}-\frac{2\pi}{c'^{2}}].
\end{equation}

Eq.(80) shows explicitly the quantum structure of GBEC, being parameterized by a wavelength $\lambda$ so that, in the isotropic term, 
the wavelength appears in the denominator whereas the anisotropic term is in the numerator, which generates two cutoffs of 
isotropic (speed of light) and anisotropic (minimum speed) origin by preventing the closure of event horizon in the region of phase 
transition. In such a region, as 
$[dv/dr]_{aniso}\rightarrow\infty$, we obtain a high frequency of vibration of a proof particle, and thus the anisotropy is fundamental. 
Such vibrations are a more realistic profile of relativistic superfluid than usually treated in the literature with extremely hard EOS
($p=-\rho $)\cite{9}, where the speed of signal propagation does not vary. In short, we should have a decomposition of spectrum of
different vacuums inside GBEC\cite{32}.

From the viewpoint of escape speed for the case of a BH (attractive gravity), the region of event horizon would have an
escape speed being the speed of light ($c$). So nothing could escape because there is no particle faster than light. On the other hand, 
in the region of repulsive gravity (repulsive core or GBEC), we should think about the concept of input speed rather than escape speed. 
On this repellent surface of GBEC, the input speed is a minimum speed $V'$. Therefore, all could enter into the core, since nothing 
is slower than the minimum speed. In this region of phase transition where $r=R_c$, we have an abrupt variation of the gravitational
field by changing from a regime of attractive to repulsive gravity, although we still do not know exactly how this abrupt variation occurs. 
We just know that $[\frac{dv}{dr}]_{aniso}$ should be too large in this region (surface of the core, i.e., $r=R_c$).

\subsection{Study of anisotropy in the limit of $r\rightarrow 0$}

A recurring requisition in works of Astrophysics is that $\Delta(r=0)=0$, implying that $p_{t}=p_{r}$. However, here we seek to 
circumvent the singularities such as the singularity of coordinates of event horizon replaced by a phase transition and the central 
singularity. So we can make the following approximation:

\begin{equation}
r=\frac{R_{core}V'}{v}
\end{equation}

As we know that the speed $v$ has an upper limit given by the speed of light ($c'$) inside the GBEC, the other three values such as 
$r$, $R_c$ and $V'$ are non-zero constants. Thus, if $v=c'$, we obtain a minimum radius associated with a minimum anisotropy 
($\Delta\approx 0$) so that we have $p_r\approx p_t$. This minimum radius is given by

\begin{equation}
r_{min}=R_{core}\xi',
\end{equation}
where we should emphasize that this minimum radius should be interpreted as the most inner bubble of the vacuum core (GBEC), which does not
mean that this region is inaccessible, since one can penetrate it, although the repulsion force grows alarmingly for $r\rightarrow 0$.

Recalling that the pressure $p_r$ and $p_t$ are given by 

\begin{equation}
p_{r}(r)=-(\frac{v(r)}{c'})^{2}\rho,~ p_{t}(r)=\frac{v^{2}(r)}{4{\pi}r}\frac{d^{2}m(r)}{dr^{2}}. 
\end{equation}

And the anisotropy is given by 

\begin{equation}
\Delta(v{\approx}V')=\frac{3}{4}\frac{\frac{V'^{4}}{v^{4}(r)}}{(1-\frac{V'^{2}}{v^{2}(r)})^{\frac{5}{2}}}=
\frac{3}{4}\frac{(\frac{r}{R_{core}})^{4}}{(1-(\frac{r}{R_{core}})^{2})^{\frac{5}{2}}}=\frac{p_{t}(r)-p_{r}(r)}{\rho_{eff}}.
\end{equation}

Then by substituting Eq.(83) into Eq.(84), we write

\begin{equation}
\Delta(v{\approx}V')=\frac{3}{4}\frac{\frac{V'^{4}}{v^{4}(r)}}{(1-\frac{V'^{2}}{v^{2}(r)})^{\frac{5}{2}}}=
\frac{3}{4}\frac{(\frac{r}{R_{core}})^{4}}{[1-(\frac{r}{R_{core}})^{2}]^{\frac{5}{2}}}=
\frac{\frac{v^{2}(r)}{4{\pi}r}\frac{d^{2}m(r)}{dr^{2}}+\frac{v^{2}(r)\rho_{eff}}{c'^{2}}}{\rho_{eff}}.
\end{equation}

If we consider $\Delta=0$ with $\rho$ being constant, we recover the isotropic case and thus also the continuity equation, that is 

\begin{equation}
\frac{dm(r)}{dr}={\pi}r^{2}\frac{\rho}{c'^{2}}. 
\end{equation}

Therefore we have $p_{r}=p_{t}$ for the isotropic case. However, by rewriting the second derivative of mass with respect to $v$,
we obtain

\begin{equation}
\frac{v^{2}(r)}{4{\pi}r}\frac{d^{2}m(r)}{dr}=
{\rho_{eff}}(\frac{3}{2}\frac{\frac{V'^{4}}{v^{4}(r)}}{(1-\frac{V'^{2}}{v^{2}(r)})^{\frac{5}{2}}}-\frac{v^{2}(r)}{c'^{2}}). 
\end{equation}

In Eq.(87), by taking the limit of $r\rightarrow r_{min}=R_{core}\xi'$, we obtain 

\begin{equation}
\frac{c'V'}{2{\pi}R_{core}}\frac{d^{2}m}{dr^2}={\rho_{eff}}(\frac{3}{2}\frac{\xi'^{4}}{(1-\xi'^{2})^{\frac{5}{2}}}-1)
\end{equation} 

We must stress that the minimum speed may vary in each gravastar (GBEC), where $\xi'=\frac{V'}{c'}$ is the ratio of these two speeds for
a given GBEC. And it is known that many authors already detected variation of the speed of light in a scenario of condensed
matter\cite{25}, as well as astronomical observations. This topic is still a subject of intense debate in 
the community\cite{26}\cite{27}\cite{28}\cite{29}\cite{30}\cite{31}. 

In Eq.(88), we see that we have $p_{t}{\approx}p_{r}$ in the center of a gravastar. However, such pressures are not exactly equal and 
therefore a certain minimum anisotropy still remains close to the center of the Dark Star\cite{21}\cite{22}\cite{23}\cite{24}. Here we take into 
account that we are dealing with a quasi-collapsed object, which is thanks to the existence of a minimum speed, so that we finally have
also shown that there is no central singularity, i.e., according to the conception of spacetime in SSR, the elimination of the singularity
of horizon by means of a deeper justification of the origin of phase transition also implies in the elimination of the central
singularity. Of course, we still need an improvement of the present model in order to investigate how the minimum speed and speed of light
vary close to the region of phase transition. We expect that the values of both speeds become too close to each other for
$r\approx R_{core}$, so that a strong gravity and anti-gravity could coexist in such a region. This effect justifies better why
there is no formation of event horizon. Such a question will be introduced in the next section. 

\section{Region of coexistence between phases}

Rewritting the Schwarzschild metric, we have 

\begin{equation}
ds^{2}=-c'^{2}dt^{2}+\frac{dr^{2}}{\left(1-\frac{2Gm}{c^{2}r}\right)}+r^{2}d\Omega^2=
-c^{2}\left(1-\frac{2Gm}{c^2 r}\right)dt^{2}+\frac{dr^{2}}{\left(1-\frac{2Gm}{c^{2}r}\right)}+
r^{2}(d\theta^2+\sin\theta^2d\phi^2),
\end{equation}
from where we obtain an effective speed of light $c'=c\sqrt{1-2Gm/c^2r}<c$. This assumption will be better justified soon when
we consider the light cone in the region of phase transition or close to the event horizon in the case of classical collapse\cite{18}. 

We expect that, in the region of coexistence between the two phases (gravity or baryonic phase/anti-gravity or GBEC), i.e., for 
$R_{coexistence}\approx R_{core}=R_c$, the light cone becomes closed\cite{18} so that the speed of light is reduced to 
$c'=c(R_c)=c\sqrt{1-2Gm/c^2R_c}$, leading to the closing of the light cone. However, thanks to the minimum speed, i.e., the presence of the 
dark cone (Fig.1)\cite{3}, here we intend to show more clearly that the horizon is almost formed at the Schwarzschild radius ($R_S=2Gm/c^2$),
but it is not exactly formed due to the appearence of a coexistence radius of phases ($R_c$), $R_c$ being slightly larger than $R_S$ such
that the horizon becomes forbidden.

In order to obtain the coexistence radius, we have to admit that the minimum speed increases close to the region of phase transition 
in such a way that $V'$ approaches to the speed of light $c'$ that decreases so that the cone becomes almost closed, but not exactly closed
at horizon as occurs in the classical gravitational collapse\cite{18}, by preventing the singularity of event horizon. 

To know how the minimum speed increases, we have to take into account the concept of reciprocal velocity\cite{5}, where we have seen that
the minimum speed works like a kind of ``inverse'' (reciprocal) speed ($v_{rec}$) of the speed of light, i.e., 
$v_{rec}=cV/v=v_0^2/v$\cite{5}, ($\psi(v_0)=\psi(\sqrt{cV})=1$ (Eq.4)), such that we find $v_{rec}=V$ for $v=c$. Thus, according to this 
relation for $v_{rec}$, we can get the effective minimum speed $V'$ in the region of phase transition, namely 
$V'(R_c)=cV/c(R_c)=cV/c\sqrt{1-2Gm/c^2R_c}=V/\sqrt{1-2Gm/c^2R_c}$, where $V$ is the universal (cosmological) minimum speed\cite{3}. As
$V'$ approaches to $c'$ close to the region of phase transition, both speeds become equal at $R_c$, i.e., $V'(R_c)=c'(R_c)$ so that we 
write 

\begin{equation}
c'=c\sqrt{1-\frac{2Gm}{c^2R_c}}=V'=\frac{V}{\sqrt{1-\frac{2Gm}{c^2R_c}}},
\end{equation}
from where we obtain

\begin{equation}
 R_c=\frac{2Gm}{c^2(1-\xi)}=\frac{2Gm}{c^2\left(1-\frac{V}{c}\right)}, 
\end{equation}
where $\xi$ is the universal constant of fine adjustment\cite{3}. Here we have considered that the coexistence radius of phases is practically 
equal to the core radius ($R_c$). 

Eq.(91) shows the expected result by indicating that the event horizon is not formed, since now we can see that the radius $R_c$, where the
phase transition occurs is in fact slightly larger than the Schwarzschild radius ($R_S=2Gm/c^2)$ due to the universal minimum speed
$V\sim 10^{-14}$m/s and also the universal tiny constant $\xi=V/c\sim 10^{-22}$\cite{3}, i.e., we find $R_S/R_c=(1-\xi)$. Therefore, indeed we 
conclude that the universal minimum speed is responsible for preventing the appearence of $R_S$, since during the gravitational collapse,
the phase transition from gravity to anti-gravity occurs at $R_c$ just before reaching $R_S$, that is to say $R_c>R_S$. But,
if $\xi=0$ ($V=0$), we recover the classical case of singularity at the Schwarzschild radius (no phase transition). 

As the Schwarzschild radius is not reached during the collapse in spacetime with a minimum speed, so instead of $R_S$, a mimimum radius 
for baryonic matter is reached in the region of phase transition, $R_c(=R_S/(1-V/c))$ being such a radius where the attractive matter 
is replaced by a repulsive vacuum energy of GBEC. In view of this quantum gravity effect given by the miminum speed connected to the 
cosmological constant $\Lambda$\cite{3}, we realize that the metric in Eq.(89) cannot diverge for the baryonic phase of the star since 
its minimum radius (of matter) is now $R_c>R_S$, so that the divergence of the Schwarzschild metric (Eq.(89)) vanishes. Thus, the 
divergence of the metric at $R_S$ is replaced by a too high value, being still finite. In order to obtain such a finite result, we just
substitute Eq.(91) into Eq.(89), and so we find 

\begin{equation}
ds^{2}_{max}=-\xi c^{2}dt^{2}+\frac{1}{\xi}dr^{2}+r^{2}d\Omega^{2}=-v_0^2dt^{2}+\frac{c}{V}dr^{2}+r^{2}d\Omega^{2},
\end{equation}
where $1/\xi=c/V\sim 10^{22}$ is a so large pure number, and $v_0=\sqrt{Vc}$ represents a universal speed that provides the
transition from gravity to anti-gravity in the cosmological scenario (see the last section of ref.\cite{3}). So it is interesting to note
that such a speed $v_0$ also plays the role of an order parameter obtained just in the region of phase transition of a non-classical 
gravitational collapse. This connection between the cosmological scenario with anti-gravity and the phase of a repulsive core inside 
a gravastar by means of the same universal order parameter of transition given by $v_0$ seems to be a holographic aspect of spacetime. We
will explore deeply this issue elsewhere. 

 Dirac have already called attention to the importance of the well-known {\it Large Number Hypothesis} even before the obtaining of
$\xi$\cite{3}. So a given infinite greatness that appears in Physics could be removed by a more fundamental principle still unknown. 
In view of this, it is interesting to notice that the metric in Eq.(92) shows us that the tiny pure number $\xi$ in the denominator of the
spatial term $dr$ prevents its singularity and thus also prevents an interval $ds$ of pure space-like as occurs at the event horizon of
a BH, because now the cone does not become completely closed in the region of phase transition given by the metric in Eq.(92), since the 
temporal term of the metric above does not vanish ($c'(R_c)=\sqrt{\xi}c=v_0$), i.e., $\xi c^{2}dt^{2}=Vcdt^{2}\neq 0$ as 
$v_0=\sqrt{Vc}\neq 0$, which is exactly 
the order parameter of transition that indicates the begining of a new phase of anti-gravity for $r<R_c$, as also occurs in the universe for 
a given radius such that $r>R_{critical}$, from where anti-gravity governs its expansion. However, this question will be treated elsewhere. 

It is also interesting to note that a signal could be transmitted with speed $c'=v_0$ in the region of phase transition, which does
not occur at the classical horizon where the cone is completely closed so that $c'=0$ for $r=R_S$ (no signal). Actually, the speed of
signal decreases drastically inside a very short interval of radius that begins at the coexistence radius of phases with speed $c'=v_0$
and goes rapidly to the radius of core where $c'=V$, as already shown before. However, as there is a very short and complicated
transient between $R_{coexistence}$ with signal $c'=v_0$ (Eq.92) and $R_{core}$ with a minimal signal $c'=V$, here we have made a good
approximation for both radius so that $R_{coexistence}\approx R_{core}=R_c$, although the speeds of signal vary abruptally between them.
The improvement of this model in order to distinguish both radius within a more accurate way will be made elsewhere. 

We sill realize that the Buchdahl-Bondi relation for preventing the event horizon is now better justified by the constant $\xi=V/c$,
since we find

\begin{equation}
\frac{8}{9}<\frac{2Gm}{R_cc^{2}}=(1-\xi)<1, ~~ \Delta(R_c)=\frac{1}{4}\frac{(1-\xi)}{\xi}\approx\frac{1}{4\xi}>>0,
\end{equation}
where $\xi\sim 10^{-22}$ and thus the anisotropy is in fact so large at $R_c$, i.e., $\Delta(R_c)>>0$ as already expected. 

\section{Conclusions and prospects}

In this paper, we have described GBEC of a gravastar by using SSR theory\cite{3}. So we were able to describe more clearly 
the behavior of the region of phase transition, although this issue should be explored deeply in a next article. We were also able 
to show that the vacuum (superfluid) inside the core presents several states, unlike the old model
introduced by Cattoen et al\cite{9}\cite{20} who proposed a single state of vacuum for the core. However, in the present model,
we have new typical quantities of SSR such as $V ',\xi',c '$, which are associated with a certain gravastar.

We have parameterised the GBEC with a frequency which is fully analogous to the wavelength of de-Broglie in SSR\cite{5} (quantum structure 
of space), showing that the anisotropic and isotropic terms have opposite behaviors, one of them working as a cut-off for the other, 
in such a way to prevent the closing of the event horizon.

We have also seen that the anisotropy is directly connected with the star's mass distribution (classic profile), which reorders according 
to the repulsion given by the cosmological constant of the star, which is closely related to the structure of the ultra-referential in SSR 
(quantum aspect of gravity), since it seems to be the origin of repulsion (anti-gravity). So SSR confirms its vocation to 
be an anti-gravity theory, which is able to describe a superfluid even for the macroscopic case, since SSR was generated by a breaking
of the Lorentz symmetry.

SSR opens a new perspective to the paradoxes of quantum gravity and gravitational collapse in order to be better explored. We also
realize the need to postulate a new energy condition being compatible with ultra-referential $S_V$, which will be studied in a next paper.

We have the next perspectives to study the entropy of Hawking-Berkenstein, an equation of state that is less hard for describing the 
region of core, the hydrodynamic conditions of the superfluid generated by this new dark energy and discuss the relationship of  
GBEC/SSR with the layer of baryonic matter, by taking into account the state equation $p=\rho$ to describe the region of radiation 
and the equation of MIT Bag Model for baryonic matter.

We also aim to study in detail the behavior of the light and dark cone inside the core of a gravastar and also close to the center of
the dark core, including the behavior of the constants $V'$, $c'$ and $\Lambda'$ inside the star, i.e., far away from the region of phase
transition.

\end{document}